\documentclass[a4paper]{jpconf}
\usepackage{amssymb}
\begin{document}
\title{Arguments towards the construction of a matrix model groundstate}
\author{L Boulton$^{1}$, M P Garcia del Moral$^{2}$ and A Restuccia$^{2}$}

\ead{L.Boulton@hw.ac.uk; maria.garciadelmoral@uantof.cl; alvaro.restuccia@uantof.cl}
\begin{abstract}
We discuss the existence and uniqueness of wavefunctions
for inhomogenoeus boundary value problems associated to   
$x^2y^2$-type matrix model on a bounded domain of $\mathbb{R}^2$. Both properties involve a combination of the Cauchy-Kovalewski Theorem and a explicit calculations.
\end{abstract}

\section{Introduction}
In this note we sketch a technique for determining the existence and uniqueness of groundstates of supersymmetric matrix models with global symmetry, subject to Dirichlet boundary conditions. We consider one of the simplest possible benchmark models available. This allows us to illustrate the actual procedure on a concrete setting. The hamiltonian, given by (\ref{hamil}), is the canonical toy model which is often used to test spectral properties of the $D=11$ supermembrane. In the following we call it the $x^2y^2$-model. 

We believe the approach we present below is new in the context of matrix models and its scope of applicability includes a wide range of physical settings. Wavefunctions of matrix models have been investigated by means of different methods in the past. They provide a better understanding of M-theory, either from the point of view of Supermembrane Theory \cite{dwhn, hoppe, fh} or from the point of view of other matrix models \cite{ bfss}. They also play a role in  Supersymmetric Yang-Mills theories in the slow-mode regime \cite{halpern, yi, porrati, stern}. 

\section{The $x^2y^2$-model}
The $x^2y^2$-model was first introduced in \cite{dwln}. In \cite{xy} it was shown that zero is not an eigenvalue. Numerical tests carried out in  \cite{bgmr3} suggest the existence of more general Weyl-type wavefunctions. See also \cite{korcyl}. We now show analytically that the inhomogeneous boundary value problem associated to a zero eigenvalue for Dirchlet boundary conditions on a compact domain $\Omega$ has a solution which is unique. See (\ref{bvp})-(\ref{bc}). Without loss of generality we assume that $\Omega$ is a two-dimensional ball of finite radius $R>0$.

 The supersymmetric hamiltonian to be considered is
\begin{equation} \label{hamil}
H=p_x^2+p_y^2+x^2y^2+x\sigma_3+y\sigma_1
\end{equation}
where $\sigma_i$ are the Pauli matrices. The supersymmetric charges in this case are given by the expression
\[
Q=Q^{\dag}=
\left( \begin{array}{cc} -xy & i\partial_x-\partial_y\\
i\partial_x-\partial_y & xy\end{array} \right).
\]
The wavefunctions are $\Psi=\left(\begin{array}{cc} \Psi_1 \\ \Psi_2 \end{array}\right)\in H^2(\Omega)\cap H^1_0(\Omega)$. Note they are subject to zero boundary conditions, 
\begin{equation} \label{bc}
\Psi=0\quad\textrm{on}\quad \partial\Omega.
\end{equation}

\section{Existence and uniqueness of wavefunctions at zero}
The resolvent of $H$ is compact, so the inhomogeneous system
\begin{equation} \label{bvp}
     H\Psi=\Phi
\end{equation}
for regular enough right hand side $\Phi$ will have a unique solution if and only if
zero is not an eigenvalue of $H$. We show that this is indeed the case.

Assume that 
\begin{equation} \label{zero}
     H\Psi=0.
\end{equation}
Then  
\begin{equation} \label{vanishingboth}
     Q\Psi=Q^\dag\Psi=0 \quad \textrm{in}\quad \Omega.
\end{equation}
According to the regularity properties of the elliptic operators, $H$ in this case, this condition holds true pointwise up to the boundary of $\Omega$. 

Let $\mathbf{x}\in \partial\Omega$ and denote by $(\mathbf{n}_1,\mathbf{n}_2)$ the components of the normal to $\partial\Omega$ at $\mathbf{x}$. The tangent to $\partial\Omega$ at $\mathbf{x}$ is then $(\mathbf{n}_2,-\mathbf{n}_1)$ and we must have 
\[(\mathbf{n}_2\partial_x-\mathbf{n}_1\partial_y)\Psi(\mathbf{x})=0.\]

The solution of problem (\ref{zero}) is regular, so we can extend it continuously up to the boundary. Then (\ref{vanishingboth}) yields 
\[(i\partial_x+\partial_y)\Psi_2(\mathbf{x})=(i\partial_x-\partial_y)\Psi_1(\mathbf{x})=0\]
pointwise. Since $(\mathbf{n}_1,\mathbf{n}_2)\neq 0$, if $\mathbf{n}_2\ne 0$, 
\[ 
\left(1+i\frac{\mathbf{n}_1}{\mathbf{n}_2}\right)\partial_y\Psi_2(\mathbf{x})=0 \quad \mathrm{implies}\quad \partial_y\Psi_2(\mathbf{x})=0\textrm{ and }\partial_x\Psi_2(\mathbf{x})=0,\]
and 
\[\left(-1+i\frac{\mathbf{n}_1}{\mathbf{n}_2}\right)\partial_y\Psi_1(\mathbf{x})=0 \quad \mathrm{implies}\quad \partial_y\Psi_1(\mathbf{x})=0\textrm{ and } \partial_x\Psi_1(\mathbf{x})=0.
\]
A similar conclusion is obtained for $\mathbf{n}_1\ne 0$. Hence $\Psi$ and $\partial_{\mathbf{n}}\Psi$ must vanish on the boundary $\partial\Omega$. 
 
Now, by virtue of the Cauchy-Kovalewski Theorem, cf. \cite{Folland}, and from the fact that the potential is analytic, we conclude that $\Psi=0$  pointwise on the whole of $\overline{\Omega}$.  That is, the only solution to (\ref{zero}) is the zero wavefunction.

In a paper soon to be completed \cite{bgmr4}, we will examine the existence and uniqueness of the solution of boundary value problems associated to the $SU(N)$ regularized model of the $D=11$ supermembrane on a bounded domain. This theory was introduced in \cite{dwhn} for an  unbounded domain. We expect to being able to examine the existence of massless groundstates in supersymmetric matrix models with global and gauge symmetries.

\section{Discussion} The  $x^2y^2$-model \cite{dwln} is a well-known benchmark for testing spectral properties of the $D=11$ supermembrane. The existence of a groundstate was analyzed in \cite{xy} by considering the whole space $\mathbb{R}^2$. In the latter case the model does not admit a solution \cite{xy}. We have considered this system  confined to a ball of finite radius. In this case zero is also not an eigenvalue for the space of wavefunctions which \underline{are zero} on the boundary. Additionally we show that the inhomogeneous boundary value problem associated to zero eigenvalue has always a solution which is unique. We consider a viewpoint to determine the existence of groundstate wavefunctions which we believe is new in the context of matrix models. The conditions of supersymmetry and ellipticity of the hamiltonian operator are essential to this analysis. The results announced in this note can be further generalized to more interesting settings. A rigorous general framework in this respect is currently being consider \cite{bgmr4}.

\end{document}